\documentstyle[11pt,amssymb,amsfonts]{article}
\newcommand{\be}{\begin{equation}}
\newcommand{\ee}{\end{equation}} 
\newcommand{\bea}{\begin{eqnarray}}
\newcommand{\eea}{\end{eqnarray}}

\begin{document}

\begin{titlepage}

\begin{flushright} 
{\tt 	FTUV/97-25\\ 
	IFIC/97-25\\ }
 \end{flushright}

\bigskip%\vskip3mm

\begin{center}

 {\bf  {\LARGE Canonical Equivalence of a Generic
  2D Dilaton 
  
  Gravity 
    Model
     and 
 a Bosonic String Theory}}
\footnote{Work partially supported by the 
{\it Comisi\'on Interministerial de Ciencia y Tecnolog\'{\i}a}\/ 
and {\it DGICYT}}

\bigskip 

 J.~Cruz \footnote{\sc cruz@lie.uv.es}
and J.~Navarro-Salas \footnote{\sc jnavarro@lie.uv.es}.

\end{center}

\bigskip% 

\footnotesize
	 Departamento de F\'{\i}sica Te\'orica and 
	IFIC, Centro Mixto Universidad de Valencia-CSIC.
	Facultad de F\'{\i}sica, Universidad de Valencia,	
        Burjassot-46100, Valencia, Spain.

\normalsize 

\bigskip%\vskip2mm 
%\centerline{\today}
\bigskip%\vskip2mm

%\newpage 

\begin{center}
			{\bf Abstract}
\end{center}
We show that a canonical transformation converts, up to a boundary term,
a generic 2D dilaton gravity model into a bosonic string theory with a Minkowskian target space.
 \newline
 \newline
 PACS number(s): 04.60.Kz, 04.60.Ds
% \newline
 %Keywords: Canonical Transformations, 2D gravity, string theory.
 \end{titlepage}
 \newpage
 The interest of studying two-dimensional theories of gravity has been growing in the last years.
 The main motivation is to study quantum gravitational effects in a more simplified setting. 
 In two dimensions the Einstein tensor vanishes identically and a natural analogue of the Einstein
 equations is given by the constant curvature equation. This equation can 
 be obtained from a local action \cite{Jackiw} 
 if a scalar field $\phi$ is introduced in the theory
 \be
 S=\int d^2x\sqrt{-g}\left(R+4\lambda^2\right)\phi
 \>.\label{i}
 \ee
 The constant curvature equation can also be derived from the non-local Polyakov action \cite{Polyakov},
 which can be converted into a local one by introducing a scalar field $\phi$
 \be
 S=\int d^2x\sqrt{-g}\left( 2R\phi+\left(\nabla\phi\right)^2+4\lambda^2\right)
 \>.\label{iii}
 \ee
 More recently, the model introduced by Callan, Giddings, Harvey and Strominger (CGHS) \cite{CGHS}
  \be
  S=\int d^2x\sqrt{-g}\left[e^{-2\phi}\left(R+4\left(\nabla\phi\right)^2+4\lambda^2
 \right)-{1\over2}\left(\nabla f\right)^2\right]
 \>,\label{iv}
 \ee
 where $f$ is a massless scalar field, has been extensively studied 
 because it describes the formation and evaporation of black holes in a simple way \cite{Strominger}.
 The gravitational part of the action (\ref{iv}) can be seen as a simplification
 of the spherically reduced Hilbert-Einstein action
 in four-dimensions
 \cite{Strominger}
\be
S=\int d^2 x \sqrt{-g}e^{-2\phi}\left(R+2\left(\nabla\phi\right)^2+2\lambda^2e^{2\phi}
\right)
 \>,\label{v}
 \ee
 where the 4D metric $ds^2_4$ is related to the 2D metric $ds^2$ by
 \be
 ds_4^2=ds^2+{e^{-2\phi}\over \lambda^2}d\Omega^2
 \>.\label{vi}
 \ee
  All the above models are particular cases of a large class of dilaton-gravity
  theories considered in \cite{Banks}.
It is well-known that, by a conformal reparametrization of the fields,
  one can eliminate the kinetic term for the dilaton and rewrite
  the action in the form \cite{Banks,Louis}
   \be
   S=\int d^2 x\sqrt{-g}\left(R\phi+V\left(\phi\right)\right)
   \>,\label{viii}
   \ee
   where $V\left(\phi\right)$ is an arbitrary function of the scalar field.
   For the Jackiw-Teitelboim model (\ref{i}) we have $V=4\lambda^2\phi$,
    while $V=2\lambda^2e^{-2\phi}$
   for the induced gravity (\ref{iii}), $V=4\lambda^2$ for the CGHS model (\ref{iv})
   and $V={2\lambda^2\over \sqrt{\phi}}$ for the spherically reduced Einstein gravity
   (\ref{v}).
   
   In \cite{Cangemi} it was shown that a non-local canonical transformation
    converts the constraints of the CGHS theory
    into those of a bosonic string theory with a Minkowskian target space.
    So, the hamiltonian and momentum constraint functions read as follows
    \be
    H={1\over2}\left(\pi_0^2+\left(r^{0\prime}\right)^2\right)
    -{1\over2}\left(\pi_1^2+\left(r^{1\prime}\right)^2\right)
    \>,\label{ix}
    \ee
    \be
 P=\pi_0r^{0\prime}+\pi_1r^{1\prime}
 \>,\label{x}
 \ee
 and the corresponding quantum constraints 
 $\hat C_{\pm}=\pm{1\over2}\left(\hat H\pm\hat P\right)$ generate an anomalous algebra
 \bea
 \left[\hat C_{\pm}\left(x\right),\hat C_{\pm}\left(\tilde x\right)\right]&=&
 i\left(\hat C_{\pm}\left(x\right)+\hat C_{\pm}\left(\tilde x\right)\right)
 \delta^{\prime}\left(x-\tilde x\right)
 \nonumber\\
 &&\mp{i\over24\pi}\left(c_0+c_1\right)\delta^{\prime\prime\prime}
\left(x-\tilde x\right)
 \>.\label{xi}
\eea
    It was also pointed out in \cite{Cangemi} that the
    value of the central charge $c=c_0+c_1$ depends
    on how the vacuum is defined.
    The positively signed gravity variable contributes with $c_0=1$ but the negatively signed one gives 
    a negative contribution $c_1=-1$ in the Schrodinger representation \cite{Cangemi}.
    Therefore, both contributions cancel and the 
    theory can be quantized without obstruction.
    In fact, in terms of the geometrical variables, or in the gauge-theoretical formulation
    \cite{nueva}, explicit expressions
    for the wave functions have been obtained \cite{Louis,Cangemi2}.
    However solutions to the quantum constraints of pure gravity were obtained
    in \cite{Louis} for a generic
    model of 2D dilaton gravity.
    This suggests that the equivalence of the CGHS model
    and
    a conformal theory of two free scalar fields with opposite contributions to the hamiltonian constraint could also
    be valid for a general dilaton gravity theory. 
   
    In a recent work \cite{Cruz}
    a first step in this direction was done by constructing
     explicit canonical transformations
    which convert the Jackiw-Teitelboim model 
   and the model with an exponential (Liouville)
     potential
    into a bosonic string theory with a Minkowskian target space. 
    In this paper we shall extend the results of Ref \cite{Cruz} for a general model of 2D dilaton gravity.
    We shall demonstrate the existence of a canonical transformation mapping, up to a boundary term,
    a generic model of 2D dilaton gravity into a bosonic string theory with a flat target space 
    of indefinite signature.
    
    Let us consider the action (\ref{viii})
    minimally coupled to a massless scalar field $f$. Parametrizing the two-dimensional
    metric as \cite{Louis}
     \be
  g_{\mu\nu}=e^{2\rho}\left(\begin{array}{cc}v^2-u^2&v\\v&1\end{array}\right)\>,\label{xii}
  \ee
  the hamiltonian form of the action read as
   \be
  S=\int d^2x\left(\pi_{\rho}\dot\rho+\pi_{\phi}\dot\phi+\pi_f\dot f-uH-vP\right)\>,
  \label{xiii}
  \ee 
  where the constraint functions $H$ and $P$ are given by
    \be
 H=-{1\over2}\pi_{\rho}\pi_{\phi}+2\left(\phi^{\prime\prime}-\phi^{\prime}\rho^{\prime}
 \right)-e^{2\rho}V\left(\phi\right)+{1\over2} 
 \left(\pi_{f}^2+f^{\prime 2}\right)\>,\label{xiv}
 \ee
 \be
 P=\rho^{\prime}\pi_{\rho}-\pi_{\rho}^{\prime}
 +\phi^{\prime}\pi_{\phi}+\pi_{f}f^{\prime}\>.\label{xv}
 \ee
Our strategy to prove the existence of a canonical transformation converting an arbitrary
2D dilaton-gravity model into a bosonic string theory will follow the procedure used in \cite{Cruz}.
In doing so we should find the general solution to the equations of motion of the model in conformal gauge
  $ds^2=-e^{2\rho}dx^+dx^-$,
 \be
 8e^{-2\rho}\partial_+\partial_-\rho=-V^{\prime}\left(\phi\right)\>,\label{xvi}
 \ee
 \be
 -4e^{-2\rho}\partial_+\partial_-\phi=V\left(\phi\right)\>,\label{xvii}
 \ee
 \be
 \partial_{\pm}^2\phi-2\partial_{\pm}\phi\partial_{\pm}\rho=T_{\pm\pm}^f={1\over2}\left(\partial_{\pm}f\right)^2
 \>,\label{xviii}
 \ee
 in terms of four arbitrary chiral functions and employ it to construct a canonical transformation
 mapping the theory into a parametrized scalar field theory \cite{Kuchar}.
 However a general solution to this system of equations remains elusive, so we shall first consider 
 the situation when one chiral sector of matter is trivial.
 In this case one can explicitly relate the fields $\rho$, $\phi$ and
 $\partial_{\pm}\phi$  with two chiral functions.
 This relation turns out to be sufficient to show the existence
 of a 
 canonical transformation which converts 
 the constraint functions (\ref{xiv}-\ref{xv}) into those of a parametrized
 chiral scalar field theory.
 This result can be extended immediately to the general situation,
 without imposing any restriction to the matter energy-momentum tensor.
 Finally, a canonical transformation relating a parametrized scalar
field theory to a bosonic string theory with a Minkowskian target space
 will complete the proof. 
 
 Therefore, let us start our analysis considering a
 generic theory with the restriction 
 $T^f_{--}=0$.
 It is not difficult to check that the solution to the equations
 (\ref{xvi}-\ref{xviii}) can be alternatively expressed as the solution to the equations
 \be
 e^{-2\rho}\partial_-\phi=a\>,\label{xx}
 \ee
 \be
 \partial_+\phi=A-{1\over 4a}J\left(\phi\right)\>,\label{xxi}
  \ee
  where ${dJ\left(\phi\right)\over d\phi}=V\left(\phi\right)$.
  The functions $A,a$ are related to the non-trivial component of the energy momentum
  tensor in the following way
 \be
 T^f_{++}=\partial_+A+{A\over a}\partial_+ a
 \>.\label{xxii}
 \ee
 Equation (\ref{xxi}) defines $\phi$ implicitly as a functional $\phi=\phi\left(A,a,\beta\right)$
 where $\beta$ is a function of the $x^-$ coordinate which appears as a constant of integration.
 We introduce now a definition which will be useful in the following.
 The symbol  $\ \widetilde{}\ $  affecting any functional of 
 the chiral functions $A,a,\beta$ means that 
 they are converted into $\tilde A,\tilde a,\tilde \beta$ which are now
 arbitrary (not chiral) functions and that the possible derivatives 
 or integrations have been replaced according to the rule
 ${\partial_{\pm}}\longrightarrow \pm \partial_x\ \left(\partial_{\pm}^{-1}
 \longrightarrow \pm\partial_x^{-1}\right)$.
 Taking into account that the dependence of 
 $\phi$ on $\beta$ must be ultralocal it is
 straightforward to prove that $\left(\tilde\phi\right)^
 {\prime}=\widetilde{ \phi^{\prime}}$
 and 
$\left(\widetilde{\partial_-\phi}\right)^{\prime}=\widetilde{\left(\partial_-\phi\right)^{\prime}}$.
% \be
%\left(\tilde \phi\right)^{\prime}=\widetilde {\partial_+\phi}
 % -\widetilde{\partial_-
 %\phi} =
 %\tilde A-{1\over4 \tilde a}J\left(\tilde \phi\right)-
 %\widetilde{\partial_-\phi} 
 %\>,\label{xxvii}
 %\ee
 %\be
 %\left(\widetilde{\partial_-\phi}\right)^{\prime}=
 %\widetilde{\partial_+\partial_-\phi}
 %-\widetilde{\partial_-^2\phi} =-{1\over4 \tilde a}V\left(\tilde \phi\right)
% \widetilde{\partial_-\phi} -
 %\widetilde{\partial_-^2\phi}
 %\>.\label{xxviii}
 %\ee
Following the lines of \cite{Cruz} we consider now a transformation to the new set of variables $\tilde A,
 \tilde a,\tilde b$
 \be
  \phi=\tilde\phi\>,\label{xxiii}
  \ee
  \be
  \pi_{\phi}=-2\widetilde
 { \dot\rho}=
  {1\over4 \tilde a}V\left(\tilde\phi\right)+{\tilde a^{\prime}\over \tilde a}
  -\widetilde{\left({\partial^2_-\phi\over\partial_-\phi}\right)}
  \>,\label{xxiv}
  \ee
  \be
  \rho={1\over2}\log {\widetilde{\partial_-\phi}\over \tilde a}\>,\label{xxv}
  \ee
  \be
  \pi_{\rho}=-2\widetilde{\dot \phi}=
  -2\tilde A +{1\over2 \tilde a}J\left(\tilde\phi\right)
  -2\widetilde{\partial_-\phi}\>,\label{xxvi}
  \ee
  After a long computation the symplectic 2-form
  on the unconstrained phase space
  \be
  \omega=\int dx\left(\delta\phi\wedge\delta\pi_{\phi}+
  \delta\rho\wedge\delta\pi_{\rho}+\delta f \wedge\delta\pi_f\right) 
  \>, \label{xxvii}
  \ee
  becomes (from now on the exterior product will be omited)
  \be
  \omega=\int dx\left( 2{\delta\tilde a\over \tilde a}\delta\tilde A 
  +\delta f\delta\pi_f\right)+\omega_b\>,\label{xxviii}
  \ee
  where $\omega_b$ is a boundary term
  \be
  \omega_b=\int d\left(\delta\tilde\phi{\delta\tilde a\over\tilde a}+\delta\tilde\phi{\delta
  \widetilde{\partial_-
  \phi } \over \widetilde{\partial_-\phi}}
   \right)\>,\label{xxix}
  \ee
  and the light-cone combinations $C^{\pm}=\pm{1\over2}\left(H\pm P\right)$ of the constraints
  (\ref{xiv}-\ref{xv}) turn out to be
  \be
  C^{+}=\tilde A^{\prime}+\tilde A{\tilde a^{\prime}\over \tilde a}+{1\over4}
  \left(\pi_f+f^{\prime}\right)^2\label{xxx}
  \ee
  \be
  C^-=0\>.\label{xxxi}
  \ee
  Equation (\ref{xxxi}) is consistent with the assumption of chirality $\left(T^f_{--}=0\right)$.
  At this point,
  it is clear that
  defining 
  \be
  X^+=\log \tilde a \tilde A,\qquad
   \Pi_+=2\tilde A\>,\label{xxxii}
   \ee
  the 2-form (\ref{xxviii}) becomes
  \be
  \omega=\int dx\left(\delta X^+\delta \Pi_++\delta f\delta \pi_f\right)+\omega_b
  \>,\label{xxxiv}
  \ee
  and, therefore, $\left(X^+,\Pi_+\right)$ become canonical coordinates
  for the gravity-sector up to a boundary term $\omega_b$.
  Moreover the constraints
  (\ref{xxx}-\ref{xxxi}) are now the constraints of a parametrized
  chiral scalar field theory
  \be
  C^+=\Pi_+X^{+\prime}+{1\over4}\left(\pi_f+f^{\prime}\right)^2\>,\label{xxxv}
  \ee
  \be
  C^-=0\>.\label{xxxvi}
  \ee
  
  We shall now show that the canonical equivalence of a chiral 2D dilaton gravity
  model and a parametrized chiral scalar field theory can be extended to the
  non-chiral situation.
  In the general case, the 
  classical solutions are parametrized by
   four arbitrary chiral functions $A\left(x^+\right),a\left(x^+\right),
 B\left(x^-\right)$, 
 $b\left(x^-\right)$.
  We can choose $A,a,B,b$ in such a way that when $T^f_{--}=0$
 the equations of motion are equivalent to (\ref{xx}-\ref{xxi})
 and when $T^f_{++}=0$ they are equivalent to
 \be
 \partial_-\phi=B-{1\over4 b}J\left(\phi\right)\>,\label{xxxvii}
 \ee
 \be
 e^{-2\rho}\partial_+\phi=b
 \>.\label{xxxviii}
 \ee
 The classical solution $\phi=\phi\left(A,a;B,b\right)$,
  $\rho=\rho\left(A,a;B,b\right)$ defines a transformation:
  $\phi=\tilde \phi,\ \pi_{\phi}=-2\widetilde{\dot\rho},
\ \rho=\tilde\rho,\ \pi_{\rho}=-2\widetilde{\dot\phi}$,
 which reduces to (\ref{xxiii}-\ref{xxvi}) for $B=0$ and to the analogous chiral transformation for $A=0$.
 It is then clear that the only possible form for the constraint functions consistent with the
 previous result is
  \be
 C^{\pm}=\Pi_{\pm}X^{\pm\prime}\pm{1\over4}\left(\pi_f\pm f^{\prime}\right)^2
 \>,\label{xxxxi}
 \ee
 where $X^+,\Pi_+$ are given by (\ref{xxxii}) and 
$X^{-}=\log \tilde b \tilde B,\ \Pi_-=2\tilde B$.
 All we need now is to see that the transformation
 $\left(\phi,\pi_{\phi},\rho,\pi_{\rho}\right)\longrightarrow\left(X^{\pm},\Pi_{\pm}
 \right)$ is canonical up to a boundary term.
 This follows immediately because the
 unique expression for $\omega$ 
    in terms of $X^{\pm},\Pi_{\pm}$
    which leads to the 
    hamiltonian equations of motion in conformal gauge, 
    $\partial_{\mp}X^{\pm}=\partial_{\mp}\Pi_{\pm}=0$,
    and is the integral
    of a scalar density is (omiting the matter contribution and boundary
    terms)
 \be
 \omega=\int dx \left(\delta X^+\delta\Pi_++\delta X^-\delta\Pi_-\right)
 \>.\label{xxxxiv}
 \ee
 A further linear canonical transformation
\cite{Kuchar3}
 \be
 2\Pi_{\pm}=-\left(\pi_0+\pi_1\right)\mp\left(r^{0\prime}-r^{1\prime}\right)
 \>,\label{xxxxv}
 \ee
 \be
 2X^{\pm\prime}=\mp\left(\pi_0-\pi_1\right)-\left(r^{0\prime}+r^{1\prime}\right)
 \>,\label{xxxxvi}\ee
 converts the constraints of a parametrized
 scalar field theory into those a bosonic string theory
 with a Minkowskian target space. 
 The boundary term $\omega_b$ of the symplectic form can be treated in two different ways.
 Either impossing appropriated boundary conditions to the $X^{\pm}$ 
 fields to cancel it when the spatial section is closed \cite{Cruz}
  or introducing new variables
 to account for the asymptotic behaviour of the fields 
 when the spatial section is open \cite{Barvinski,Kuchar2}.
 In the absence of matter fields the equivalence of a dilaton gravity model
 and a set of two free fields of opposite signature explains why there is no
 obstruction in the quantization of a generic model in the functional Schrodinger
  approach \cite{Louis}.
  When matter fields are present the quantum constraints require a modification
  to cancel the anomaly. Remarkably, the addition of a $X^{\pm}$-dependent 
  term to the quantum constraints (\ref{xxxxi}) (motivated by a covariant 
  factor ordering \cite{Kuchar3}) produces a cancellation of the anomaly allowing 
  to solve all the Dirac quantum conditions \cite{Kuchar2,Benedict}
  maintaining the unitarity of the theory.
  Therefore it is possible to consistently quantize a matter-coupled dilaton gravity theory in the
  functional Schrodinger approach.
  \newline
  
  We thank J. M. Izquierdo, A. Mikovic, D. J. Navarro, M. Navarro 
  and C. F. Talavera for useful comments. 
  J.C. acknowledges the Generalitat
  Valenciana for a FPI fellowship. 
  
  \end{document}